# Comparing the Run-time Behavior of Modern PDES Engines on Alternative Hardware Architectures

Romolo Marotta
romolo.marotta@gmail.com
DICII
University of Rome "Tor Vergata"
Rome, Italy

Francesco Quaglia
francesco.quaglia@uniroma2.it
DICII
University of Rome "Tor Vergata"
Rome, Italy

## ABSTRACT

The current trend of technology has brought parallel machines equipped with multiple processors and multiple memory sockets to be available off-the-shelf—or via renting through Iaas Clouds—at reasonable costs. This has opened the possibility of natively supporting HPC in diffused realities, like industry or academic labs. At the same time, the Parallel Discrete Event Simulation (PDES) area has given rise to attractive simulation engines, designed with orientation to high performance and scalability, also targeting differentiated exploitation of the specific support offered by the underlying hardware. In this article, we present an experimental study where we deploy two last-generation open-source PDES platforms—one optimistic (USE) and one conservative (PARSIR)—on top of two significantly different hardware chipsets based on either x86 CISC or powerPC RISC technology, both offering multiple Non-Uniform-Memory-Access (NUMA) nodes and multiple tens of cores and hardware-threads (logical CPUs). Also, we consider real-world simulation models configured in a variety of different manners in order to investigate the actual execution profile of the PDES engines on the two distinct hardware platforms. Our objective is the one of providing insights on current performance trends, which can support decisions in terms of both strategies—for software platforms to adopt—and investments—in terms of hardware platforms—in the area of discrete event simulation.

## 1 INTRODUCTION

Hardware parallelism is nowadays a common characteristic for almost any machine, and the industrial world has made hardware platforms equipped with multiple processors—each one hosting multiple cores and multiple hardware-threads (seen as logical CPUs from the operating systems)—as well as multiple memory sockets [15] available off-the-shelf at reasonable costs. The possibility to support HPC thanks to the exploitation of such shared-memory parallel machines has therefore become a diffused reality.

At the same time, although reference protocols exist for creating hardware—like for example MESI or MOESI for supporting cache coherence [11]—the final implementation of these protocols might be significantly different in chipsets by different vendors. The same can be true for other hardware facilities, like the support for memory consistency, which may require differentiated interactions between the CPU store-buffers and the memory hierarchy (cache plus RAM), leading to differentiated impact in terms of the clock-cycles required for exposing written values (through actual memory updates) to the other CPUs. Additionally, the implementations of the aforementioned supports lead to different relative incidence on software execution also in relation to the characteristics of the supported ISA, which could be either CISC or RISC.

At the same time, the literature in the area of Parallel Discrete Event Simulation (PDES) has proposed clever PDES platforms designed considering various different aspects to be exploited for managing the workload of simulation events to be processed (e.g. [4, 5, 13, 22]). Contextually, the various implementations of these PDES platforms lead to stressing the actual protocols implemented in the hardware chipsets differently. In fact, these platforms may need more or less ample usage of atomic Read-Modify-Write (RMW) instructions for managing memory locations—we recall these instructions impact both cache coherency and the interaction between CPU store buffers and memory—and may offer more or less deep optimizations in terms of interactions with memory banks—-an aspect that is fundamental in NUMA platforms, in particular for controlling cache miss delays.

On the basis of the above considerations, in this article we focus on two last generation open-source PDES systems and deploy them over two significantly different hardware platforms—RISC vs CISC—in order to provide insights on how the performance/scalability requirements of the PDES platforms are actually supported by the different hardware chipsets. Additionally, our study also demonstrates which specific optimization that is offered by the PDES engine can be more relevant depending on the underlying chipset that is exploited for model execution.

The two PDES platforms we selected, both based on C technology, are USE (Ultimate-Share-Everything) [13] and PARSIR (PARallel SImulation Runner) [22][1]. Our choice is motivated by the fact that they are fully orthogonal, given the different approach used for synchronization—optimistic for USE and conservative for PARSIR—and the reliance on definitely different mechanisms at the level of the simulation engine—fully shared event pool for USE, managed via non-blocking algorithms requiring large incidence of RMW instructions, vs mesh-based event queues for PARSIR that mostly lead to disjoint-access parallelism. Also, these two platforms have different mechanisms for improving cache and NUMA locality. Overall, they represent fully complementary software systems for assessing differentiated hardware platforms. Additionally, our analysis also considers differentiated overlying simulation applications, which may have execution patterns that can be supported in a different manner by the underlying hardware architecture, if only because of the relevant distance between RISC and CISC ISA.

[1]The code repositories for the platforms are available at [12, 23].

As for the hardware side, in our study we consider two modern shared-memory parallel platforms. The first is based on x86 CISC technology and offers 20 processor-cores that become 40 logical CPUs (with a maximum of two hardware threads per core) working on a 2-socket setup with two different NUMA nodes. The second is based on powerPC RISC technology and relies on a 2-socket setup—where each socket is a dual chip module—offering 4 NUMA nodes and 96 logical CPUs deployed on top of 12 processor cores.

We feel our study offers an interesting analysis that is typically missing in the literature since previous works focusing on analyzing and optimizing PDES platforms, on the basis of the features of the underlying hardware, typically assess solutions on some individual hardware chipset [3, 6, 28, 29]. Hence, the concept of comparative analysis of the effects of homologous (fully CPU-based) but different hardware on the execution of PDES software platforms appears to be missing.

Overall, the contributions of this article can be summarized as follows:

- we provide a study that enables determining what kind of parallel hardware chipset can be more suitable for PDES systems offering different design choices and implementations, and when also considering specific simulation model features, like for example the more or less massive usage of arithmetic operations;
- the study is based on a combination of results coming from two significantly different hardware chipsets (RISC vs CISC) and two fully complementary PDES environments, USE and PARSIR;
- we discuss aspects related to possible choices the user can take in relation to both PDES environment and hardware chipset selection, which in turn can determine differentiated strategies and investments—in particular for what concerns the acquisition, or renting, of specific hardware—in the context of discrete event simulation.

The remainder of this article is structured as follows. In Section 2 we outline the major features of the PDES systems used, focusing on aspects that can more directly depend on the underlying hardware chipset. The two compared hardware chipsets are introduced in Section 3. The setup of the experimental study is discussed in Section 4. Its outcomes are illustrated in Section 5. Main indications from our study are proposed in Section 6. Related work is presented in Section 7. Conclusions are discussed in Section 8.

# 2 PDES PLATFORMS AND THEIR FEATURES

## 2.1 USE (Ultimate Share Everything)

USE is a PDES system that supports optimistic execution of simulation models based on checkpointing. The core objective of USE is the one of fully sharing the workload of all the simulation events to be processed across all the worker threads. The advantage is the possibility of using the computing power (the logical CPUs) by always picking pending events to process with lower timestamps across all the simulation objects of the model. This has been shown to definitely reduce the actual incidence of out-of-order event processing and rollback compared to platforms where an object is bound to a specific thread (up to the next object-to-thread re-balance operation). In fact, it prevents that any simulation object, bound to some unloaded thread of the PDES environment, optimistically runs a head too much along virtual time.

USE relies on a single fully shared event queue whose main characteristics are derived from the reference concurrent calendar queue presented in [17]. The core point is that the fully shared calendar queue does not rely on locks for managing concurrent operations. Rather, it relies on non-blocking algorithms that make large usage of RMW instructions. We recall that such instructions have a relevant impact along both: 1) the cache-coherency protocol and 2) the store-buffer management. In fact, in order to run a RMW instruction that touches a given memory location, the corresponding cache line needs to be acquired for exclusive usage by the CPU L1 cache, hence requiring the invalidation of the cache line content in other cache components—in particular when the location touched via RMW has been also accessed by other logical CPUs according to data sharing. Also, when the RMW instruction is executed, its effect is the one of reporting the modified value of the target memory location in the actual memory hierarchy (L1 cache and lower components), but this requires flushing to memory also all the still pending updates that are kept by the CPU store buffer.

At the same time, along its development, USE included some mechanisms for improving the memory locality of the operations by threads. In particular, in [19] a solution integrated in USE has been presented for leading a thread to put in place a kind of short-term binding with respect to a simulation object. The objective is the one of processing not just a single event before releasing the object, but instead a set of the object events, whose timestamps fall within a specific virtual time window—again the simulation object is prevented to optimistically run ahead too much. This leads the simulation object to raise up in the caching hierarchy, in particular in the cache components that link RAM memory to the logical CPU where it is CPU-dispatched. This information is further exploited—via apposite metadata kept by the USE engine—in order to make other threads favor picking simulation objects whose states have been recently raised up in some cache component closer to the CPU where the pick occurs.

Finally, if no recently fetched simulation object is likely present in some close cache component, the worker thread attempts to acquire, for processing its events, a simulation object whose state is placed on the local NUMA node where the thread is running. This also means that USE offers the support for hosting the state of simulation objects on specific NUMA nodes of the architecture. A cross NUMA-node migration mechanism is also included in order to re-balance the distribution of the simulation objects across the different NUMA nodes, depending on their load of events to be processed.

Overall, leaving out the specific features of the overlying discrete event simulation model, the USE platform appears to be characterized by the following points in terms of activities at the hardware level:

- massive usage of RMW instructions for the management of the fully shared event pool, hence massive incidence by these instructions on cache coherency protocols and store-buffer management;
- good cache exploitation for the actual processing of events at the simulation objects;
- good locality exploitation in NUMA chipsets.

As for the last point, USE deploys threads to the hardware clustering them on a same NUMA node. The same is true for the states of simulation objects. Hence, it moves using the logical CPUs that stand on other NUMA nodes just when already occupied nodes become fully busy. This choice is motivated primarily by the need for maintaining threads close to each other, in particular for keeping the effects on the hardware (cache-line access and invalidation) confined to closer hardware components.

## 2.2 PARSIR (PARallel SImulation Runner)

PARSIR enables running discrete event simulation models according to a conservative synchronization approach based on constant-global lookahead. It targets disjoint-access parallelism by organizing the event pools, one per each simulation object, into calendars. Each bucket in the calendar is managed via a spinlock, which is handled via RMW instructions. Hence, concurrent operations are enabled on the different buckets of the same queue associated with any individual simulation object. This exploits the typical case where each thread when extracting/inserting events in the calendars falls managing a different bucket. Consequently, the need for acquiring the exclusive usage of a cache line for managing the spinlock typically leads to low effects in terms of cross cache-line invalidations.

Exploitation of the caching system—in terms of locality of the activities—is achieved in PARSIR thanks to the reliance on batch-processing of all the events destined to a simulation object, which falls in the current simulation window. Hence, each worker thread picks a different simulation object for processing its events only after having processed all the current-window events destined to a previously picked object. In other words, when an object becomes hot since the thread has started working on its state in the current window, such state—which has been brought up in the caching hierarchy—continues to be accessed with low likelihood of being cache-replaced by other activities carried out by software.

The acquisition of a simulation object, in order to process its current-window events, is done by PARSIR in a non-blocking manner by exploiting RMW instructions. These instructions atomically increment a counter shared among all the threads, which allows each thread to determine the ID of the object to be processed. However, the impact by such RMW instructions on cache-line invalidation and on store-buffer flush is expected to be limited, given that the atomic counter increment is executed along time less frequently compared to the processing of individual simulation events. In fact, as pointed out before, once picked an object via the atomic counter manipulation, all its current-window events are processed before attempting another ID-pick operation.

PARSIR manages the memory destined to keep the chunks belonging to the state of any individual simulation object by locating it on some specific NUMA node. Hence, it adopts a NUMA-aware placement of the objects on the underlying platform. This approach is exploited also for improving NUMA-locality when threads pick the objects for processing their events. In fact, different counters are kept, each associated with the IDs of simulation objects that are located on a different NUMA node. Each thread gives higher priority to the acquisition of simulation objects (their IDs) that are located on the same NUMA node where the thread resides, thus favoring cache-miss latency and reducing traffic on the memory interconnection network. Interestingly, this NUMA optimization does not require migrating logical pages across the different NUMA nodes at run-time.

Also, at the end of each simulation window, all the threads synchronize using a barrier that is still implemented relying on RMW instructions. As for the previous discussion related to the acquisition of IDs of simulation objects, such usage of the RMW support for barriers occurs relatively infrequently compared to the actual processing of the events standing in each window.

Overall, still leaving out the specific features of the overlying discrete event simulation model, the PARSIR platform appears to be characterized by the following points in terms of activities at the hardware level:

- limited usage of RMW instructions for picking the IDs of simulation objects whose events need to be processed in the current window, and for running barriers at the end of the window; this is expected to limit the impact on cache coherency protocols and store-buffer management;
- good cache exploitation for the actual processing of events at the simulation objects, especially for non-minimal lookahead models;
- good locality exploitation in NUMA chipsets, still with the avoidance of page migration across NUMA nodes.

As for the later point, in PARSIR threads—as well as the states of simulation objects—are distributed circularly across the NUMA nodes just in order to exploit the hardware-offered scalability of memory access, especially considering the common disjoint-access parallelism that PARSIR supports. This is interesting since it offers a diametrically opposed solution compared to USE, which allows widening the aspects touched in our experimental analysis.

# 3 TESTED HARDWARE CHIPSETS

## 3.1 x86 CISC

Our CISC chipset in this experimental study is a multi-processor machine with 2 memory sockets. In particular, it is equipped with 2 intel Xeon Silver 4210R processors, each equipped with 10 cores and 20 Hardware-Threads (logical CPUs), working at 2.4GHz. The total number of available logical CPUs is therefore 40. The whole set of characteristics of this machine is shown in Table 1, where we also include data related to cache components (with L1 and L2 cache size listed per core) and the overall RAM memory, also reporting information related to the distribution on the 2 NUMA nodes associated with the two sockets of the chipset.

## 3.2 powerPC RISC

The used RISC architecture is based on two sockets. Each one is a dual chip module, and hosts two Power10 chips, for a total of 4 Power10 chips. Each of the 4 chips is recognized as an individual

Table 1: Details of the CISC platform

| | |
|---|---|
| Architecture: | x86_64 |
| CPU op-mode(s): | 32-bit, 64-bit |
| Byte Order: | Little Endian |
| Logical CPU(s): | 40 |
| On-line logical CPU(s) list: | 0-39 |
| Thread(s) per core: | 2 |
| NUMA node(s): | 2 |
| Vendor ID: | GenuineIntel |
| CPU family: | 6 |
| Model name: | Intel(R) Xeon(R) Silver 4210R CPU @ 2.40GHz |
| CPU MHz: | 1000.030 |
| L1d cache: | 32K |
| L1i cache: | 32K |
| L2 cache: | 1024K |
| L3 cache: | 14080K |
| NUMA node0 logical CPU(s): | 0,2,4,6,8,10,12,14,16,18, 20,22,24,26,28,30,32,34,36,38 |
| NUMA node1 logical CPU(s): | 1,3,5,7,9,11,13,15,17,19, 21,23,25,27,29,31,33,35,37,39 |
| NUMA node0 RAM: | 16GB |
| NUMA node1 RAM: | 16GB |

NUMA node. The total number of licensed processor cores is 12, and each of them supports 8 Hardware-Threads (logical CPUs), for a total of 96 logical CPUs in the systems. The peak speed of the processors is 4.0 GHz. The features of the hardware chipset are shown in Table 2, still reporting data related to cache components (with L1/L2 cache size aggregated across all the CPU-cores) and the NUMA architecture.

Table 2: Details of the RISC platform

| | |
|---|---|
| Architecture: | ppc64le |
| Byte Order: | Little Endian |
| Logical CPU(s): | 96 |
| On-line logical CPU(s) list: | 0-95 |
| Model name: | POWER10 (architected), altivec supported |
| Model: | 2.0 (pvr 0080 0200) |
| Thread(s) per core: | 8 |
| L1d: | 768 KiB |
| L1i: | 1.1 MiB |
| L2: | 24 MiB |
| L3: | 96 MiB |
| NUMA node(s): | 4 |
| NUMA node0 logical CPU(s): | 0-23 |
| NUMA node1 logical CPU(s): | 24-47 |
| NUMA node2 logical CPU(s): | 48-71 |
| NUMA node3 logical CPU(s): | 72-95 |
| NUMA node0 RAM: | 16GB |
| NUMA node1 RAM: | 16GB |
| NUMA node2 RAM: | 16GB |
| NUMA node3 RAM: | 16GB |

# 4 EXPERIMENTAL SETUP

As pointed out, our experimental study is aimed at determining how different hardware chipsets can support the execution of differentiated PDES platforms. However, we also need to note that the actual simulation model, which is executed by the PDES platform, can play a relevant role in the assessment. In particular its event execution latency may determine the percentage of wall-clock-time spent by each thread in simulation engine vs model specific instructions.

As for the model-specific part of the executable code, each simulation object is an independent entity not managing shared data with respect to the other simulation objects, or to the engine level software. Hence, we may expect that—thanks to the support for simulation object processing locality offered by both USE and PARSIR—the user-level code will not lead to real/important impact on cache-coherency protocols (for example invalidating cache lines across different components) and store-buffer flush operations—for making actually visible to other threads updates of the simulation object state that are carried out by the thread in charge of processing it.

Hence, it appears natural to consider that the finer the grain of simulation events the stricter is the effect of the hardware chipset on engine level execution, while the coarser the event grain the lighter such an effect.

However, there is a core aspect to keep into account in relation to cache/NUMA effectiveness, since the model-specific part of the binary code leads anyhow to better or worse performance/scalability depending on how the support—in terms of memory access delay—is offered by the hardware chipset, also considering cache miss events. In our analysis, we consider this aspect through the size of the state of a simulation object, and to the actual need for accessing to such an object state by traversing it while processing a specific event. Overall, the larger the state, the longer the event execution delay, which is the scenario where we can observe the impact of caching/NUMA effectiveness for model execution. This index is anyhow relevant for the evaluation of the PDES-engine part of software since the engines we are considering—USE and PARSIR—explicitly offer the support for both cache and NUMA exploitation while processing simulation events at the object level.

Additionally, it is also relevant to consider simulation-event execution flows where different types of operations—like for example floating point operations, or more generally arithmetic operations—require to be executed, which can lead to differentiated impact on the pipeline execution of the two differed chipset. As for this point, the two chipsets offer different degrees of sharing of a same processor core (and its base stations) to the logical CPUs it hosts, hence we can get different volumes of conflicts in the usage of the base stations by the logical CPUs on the two different hardware chipsets, which can further contribute to the comparison. However, in relation to this aspect, we decided to exclude from our study the scenario where large usage of the core pipeline is devoted to process events that at the same time do not require memory accesses. Hence, we do not focus on pure stress on the core pipeline. This exclusion is motivated by the fact that there already exist benchmarks suited for providing insights on the effectiveness of different core pipelines—and different levels of sharing of pipeline components across logical CPUs—under such kind of stress workload (see, e.g., [1]). At the same time, the incidence of simulation-engine software would be negligible in such a scenario, given the aforementioned absence (or negligible relevance) of cache/NUMA locality effects while CPU-dispatching simulation objects of generic models—at least for what we can expect by the underlying chipset on the basis of locality optimizations provided by the simulation engine.

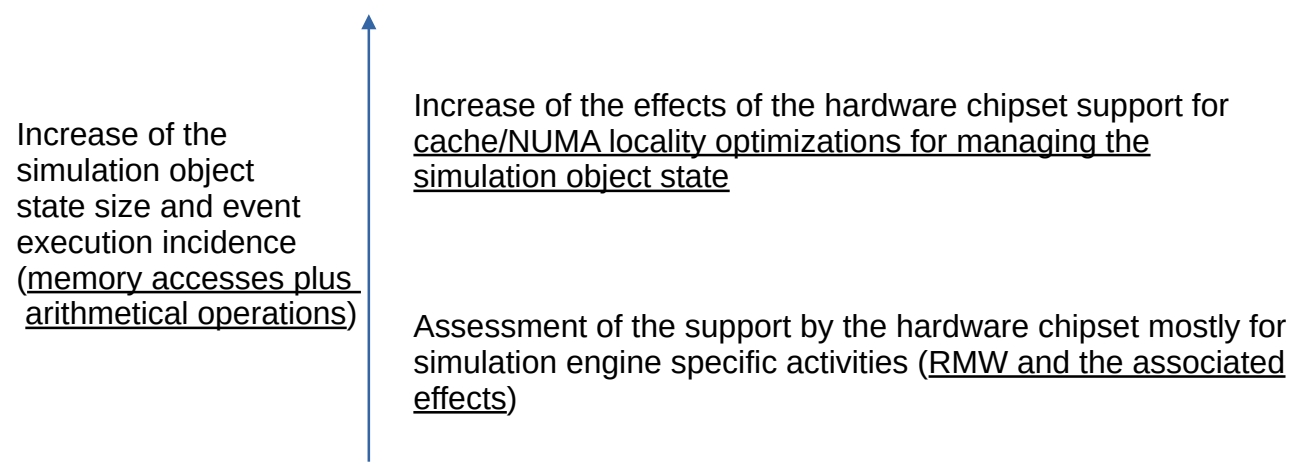

**Figure 1: Scheme of the experimental structure.**

Overall, we show in Figure 1 a concise description of the area we cover with our experimentation. Such covering has been carried out by relying on two distinct real-world models, which we describe in the following sections.

## 4.1 Personal Communication System Model

The Personal Communication System (PCS) model allows studying the evolution of a wireless communication area. Each simulation object models a communication cell that offers a given number of channels, and the workload of requests for using these channels for setting up a call determines the workload of the cell. Each time a call is started (via a proper event), one of the free channels is reserved for it, and a data structure is allocated for keeping information on the on-going call. Also, for ongoing calls, the data structure that models the channel records the communication power that has been set up. Hence, when a new call request is issued, these values are scanned and used to compute the Signal-to-Interference Ratio (SIR) that determines the amount of power to be setup for the new call, in particular according to the equations presented in [14]. We recall that this computation requires power and other arithmetical operations on `double` values loaded from memory. The length of each call is determined by an exponential distribution with mean set to 2 minutes. If the device that started the communication moves to another cell in that interval, then a handoff event occurs on the cell that is leaved, as well as on the cell that is reached. Also, in the reached cell, the setup of the incoming call is still done via a free channel, and the determination of the communication power on the basis of the evaluated SIR.

The movement of devices between cells takes place according to a random-way-point model, and the delay for migrating to a neighbor cell is determined by an exponential distribution with mean set to 5 minutes.

We simulated a square area made of 4096 cells, each one modeled as a hexagon. Hence, all the cells, except border cells of the square area, have six neighbors where handoff can take place. Each cell offers the support for 5000 channels. Hence, in our study we consider a large coverage area managing a large number of channels, which can correspond to a large city.

At the same time, as we will explain in detail in Section 5, we changed the average number of busy channels per cell, which depends on the actual frequency of communication requests that occur in each cell. This allowed us to change the granularity of simulation events—via the change of their CPU/memory-access demand (for arithmetic and data traversal operations)—which grows vs the growth of the number of active channels in a cell.

## 4.2 Highway Model

The Highway model allows studying mobility aspects. It represents different zones of a highway—in particular each different Km of the highway—via a different simulation object. The object records the exact set of cars that are currently transiting in that zone of the highway. The cars are kept by the simulation object into a list of buffers linked via pointers, which rely on dynamic memory allocation/release each time a car enters/exits the highway zone modeled by the simulation object. These cars belong to different classes which in their turn express the different cruising speeds they would reach in the scenario of non-busy lines in the highway. At the same time, the business of the lines determines a decrease in the speed of the cars in order to enable keeping the correct safe distance. The highway can be configured in order to determine access and exit points where cars can enter/exit the highway. At the same time, the initial state of the highway can be configured to have different volumes of cars in the different zones.

Each time a car enters a highway zone, the set of cars that are hosted in that zone—in particular their list—is re-accessed in order to determine whether updates in the state of the zone needs to be done. In particular, cars that move at different speeds can surpass each other, thus also determining the actual latency of traversal of the highway zone. The likelihood of occurrence of surpasses also depends on the density of cars in the zone.

In our usage of the highway model, the speed limit has been set to 130 Km per hour, as it likely occurs for various highways (motorways) all over the world. At the same time, we consider a highway made of 3000 Km (hence 3000 simulation objects), with three lines per direction.

In our usage of this model, the overall volume of cars that travel along the highway has been partitioned into three different car types, which tend to traverse the highway according to the speed limit—namely 130 Km per hour—or lower speeds—namely 120 and 110 Km per hour. In any case the actual speed the car keeps also depends on the density of cars in the specific zone they are currently traversing. More in detail, the speed of a car, beyond depending on the car type, depends on the ratio between the actual number of cars in the highway zone modeled by a simulation object and the car number limit determined by the safe distance related to the speed limit of 130 Km per hour. If this ratio is greater than one, then its value determines a statistical reduction factor for the car speed, just to ensure distance safety.

Compared to PCS, one peculiarity of the highway model is the reduced incidence of arithmetic operations in the CPU pipeline execution flow, which intrinsically leads to major incidence of memory boundness in this model. This is relevant in terms of orthogonality of this model compared to the PCS one.

Still in relation to orthogonality vs PCS, for the highway model we will consider (see Section 5) both balanced and unbalanced densities of cars in the different zones of the highway, hence leading to observe how the hardware chipsets we compare behave under balanced vs unbalanced workloads.

# 5 EXPERIMENTAL RESULTS

## 5.1 Results with PARSIR

In this section, we focus on how the compared hardware chipsets react when the PDES engine used for running the simulation models is PARSIR. We compiled it relying on the `-O3` optimization flag, thus providing the `gcc` tool with the ability to select the best combination of machine instructions in the executable for optimizing the execution speed on the specific underlying hardware architecture.

As for the performance index, we used the throughput of simulation events that are executed along a wall-clock-time second. Each reported value has been computed as the average over 20 different samples. Also, each run in our experiments has been set to last 1 minute along wall-clock-time

We configured the two models, PCS and Highway, in order to provide them with light, medium and heavy workloads. This allowed us to change the percentage of CPU usage by engine level software compared to model specific software.

In the passage from light to heavy load with PCS, we setup the model to have 120 to 1200 busy channels per cell, on the average, passing through 600 as middle load. At the same time, for the Highway model we moved the average density of cars from 0.25, to 1 and then to 1.5, the reference density that is achievable considering car speed of 130 Km per hour (the limit speed), while keeping the safety distance.

As noted, for PARSIR, larger incidence of engine level software along the execution of threads is expected to provide reduced incidence of RMW instructions, cache-line invalidation and store-buffer flushes, just given the mesh-based paradigm for the implementation of the event-pools of the different objects and the traversal of different lists associated with different buckets when inserting events. However, the engine-level software uses arithmetic operations. This takes place, in particular for managing the traversal of any bucket of the calendar associated with a simulation object resulting as the destination of a new event that gets produced along time while processing some other event. In fact, the timestamp of such an event (a `double` value in the PARSIR implementation) needs to be compared with the other timestamps of the events in the destination bucket for determining where to insert the new event in the bucket list. This gives pressure to the processor pipeline for the support of the arithmetical operations offered by the ISA.

The results—in particular, the event throughput values achieved when running with PARSIR—are shown in Figure 2 for the PCS model and in Figure 3 for the Highway model.

From the results we have a few clear indications, related to how the different hardware chipsets perform when running the PDES engine and the top standing simulation models. In particular, we observe a clear gain of the x86 CISC chipset when the relative weight of event processing is lower (light models) compared to the simulation-engine level execution of machine instructions. However, when the actual load provided by the simulation model increases, we observe an improvement of the actual performance guaranteed by the powerPC RISC chipset. The motivation for this trend is related to the deviation of the execution of software towards a more intensive CPU/memory bound profile, considering in particular the access to disjoint memory areas (simulation object states). Concerning this point, a relevant consideration in relation to the hardware chipset structure is that while in the x86 CISC chipset all the logical CPUs that are hosted on a same socket (a same processor) fully share the caching components if they are hosted on the same core, in the powerPC RISC chipset this does not occur. In particular, the 8 logical CPUs that are hosted on the same core are divided into 2 groups—each made of 4 logical CPUs—and share caching components only inside each individual group. Overall, the powerPC RISC chipset actually leads to enlargement of the per logical CPU cache storage that can be effectively exploited especially when memory is managed according to disjoint access parallelism.

Still in relation to this aspect, for light models another interesting observation is related to the behavior of the throughput on top of the powerPC RISC chipset. In particular, we have that the increase of the throughput when increasing the number of logical CPUs used for running the simulation models oscillates. This can be noted when running the models (or at least one of them) with 40 or 72 threads. With this setup, we note a decrease of the throughput, which then starts growing again. The motivation is still linked to the partial sharing of caching components across logical CPUs hosted on a same core. In fact, when passing through 40 or 72 total logical CPUs used by PARSIR, according to round-robin distribution of threads on these CPUs, a new set of caching components is involved in the management of the execution, which gives rise to increased pressure on cache-coherency protocols, while at the same time not favoring parallelism in the model execution at a reasonable level, in particular for the light simulation model load. In fact, with light load the simulation engine activities take place in a more intensive manner, leading anyhow to shared accesses that can affect cache-line invalidation. This phenomenon is not noted with the x86 CISC chipset since, starting from 20 threads, all the caching components in the architecture are already involved in the execution. Hence, the additional parallelism we have when increasing the number of threads beyond 20 is not yet counterbalanced by the involvement of such additional components and the increase of the effects of their management at the cache-coherency level.

We also note the different responses we have from the different hardware chipsets when the actual operations by threads are more or less CPU demanding compared to memory demanding. In particular, the PCS model is more CPU demanding compared to the Highway model since events tend to rely on arithmetical operations at the level of the processor pipeline in more intensive manner (i.e. for the evaluation of the SIR value and the power assignment to a newly incoming call), while the Highway model tends to most rely on memory accesses and predicates' evaluation, leading to conditional jumps for determining the evolution (e.g. surpasses) of the set of cars residing in each Km of the highway. The latter execution profile appears to be better supported by the powerPC RISC chipset, leading to improving the event throughput—compared to what happens on x86 CISC—more than what is observable with the PCS model.

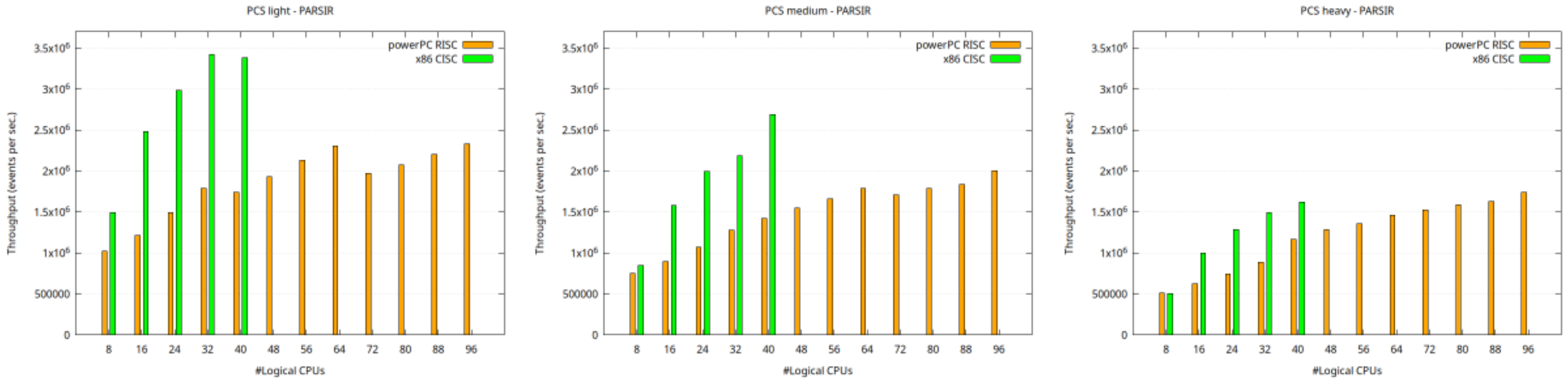

**Figure 2: Results of PCS runs on top of PARSIR.**

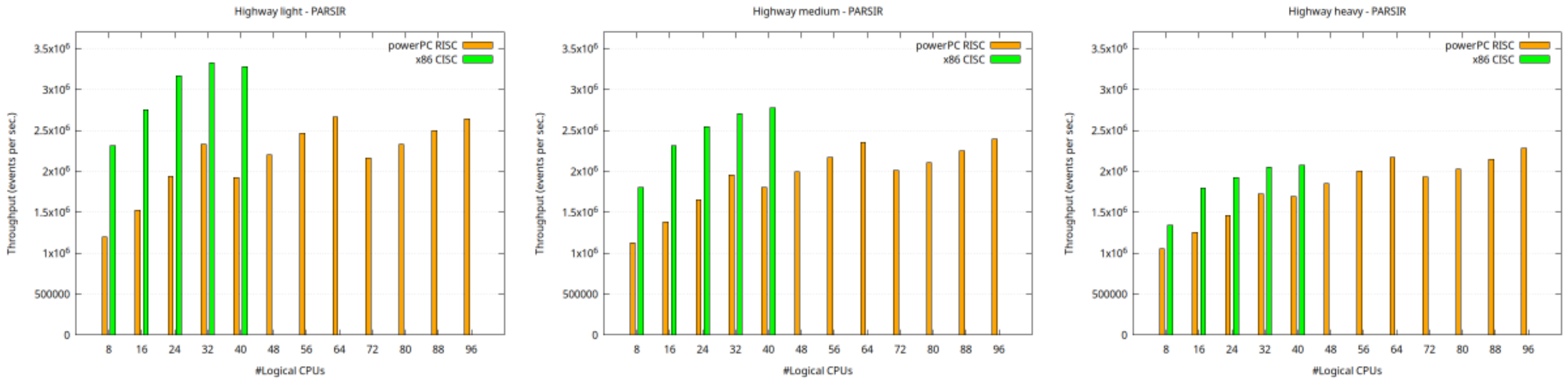

**Figure 3: Results of Highway runs on top of PARSIR.**

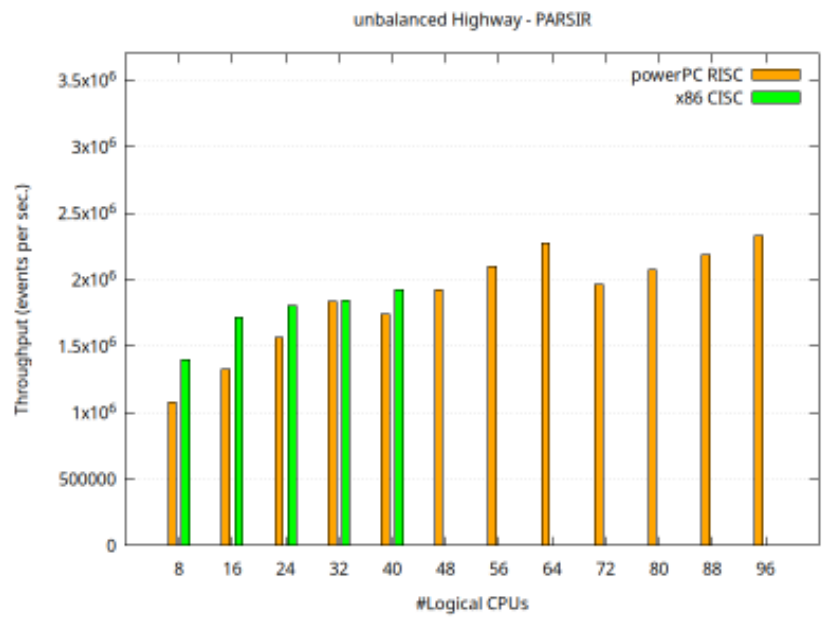

**Figure 4: Results for unbalanced Highway run on top of PARSIR.**

Still in relation to what we observed before, for the Highway model, memory boundness allows us to also observe the oscillation of the event throughput under powerPC RISC with larger number of logical CPUs and when the newly employed logical CPUs require the involvement of not yet used caching components in the architecture. Rather, for the PCS model, this phenomenon tends not to appear, in particular with larger workload, just because a larger number of CPU cycles is requested for serving instructions internally to the pipeline (rather than with interaction with memory components). As for the latter point, we also note that the two hardware chipsets have different numbers of logical CPUs hosted by an individual processor core. Hence, powerPC RISC, which offers 8 logical CPUs per core, can lead to higher volumes of conflicts in the usage of reservation stations (devices internal to the core) across the threads. This leads to increasing the number of CPU cycles required for passing through the processor pipeline, especially when arithmetic operations need to be executed by the thread in a more massive manner—this is the PCS case.

The results in Figure 2 and Figure 3 have been obtained considering a uniform distribution of the workload of events across the different simulation objects—all the cells of PCS have the same average number of calls and each Km of the Highway model has the same average number of cars in transit. In their turn, these objects are equally split among the different NUMA nodes of the hardware chipset. Given that PARSIR picks object-IDs for processing the associated events giving higher priority to the simulation objects located on the same NUMA node where the thread is running, we get that the threads essentially work in NUMA locality when accessing the object states.

In order to determine if the overall NUMA architecture embedded in the two hardware chipsets has additional different trends when memory accesses take place from remote NUMA nodes, we exploited the unbalanced configuration of the Highway model. In this configuration, half of the simulation objects (hence half of the highway) has larger load of in transit cars (equal to the one used for the balanced heavy configuration), while the other half has a load of cars reduced by 50%. In this scenario, after processing the NUMA-local objects with higher priority, PARSIR threads will start picking NUMA-remote objects, in particular the ones modeling the heavily loaded part of the highway—in fact these are the objects that other threads could have not yet been able to process in the current simulation window. This allows us to stress the response from the two chipsets when also considering the hardware level interconnection across the NUMA nodes (not just NUMA-local accesses to simulation object states). The results for this experiment are reported in Figure 4. The data show how such a scenario allows the powerPC RISC hardware to further improve its benefits compared to the x86 CISC one, hence still showing effectiveness in the hardware-level protocols for serving memory-bound execution profiles in a complex NUMA deploy, especially with the exploitation of disjoint access parallelism (for processing the events at the objects) and a lighter usage of shared data accesses at the simulation-engine level.

## 5.2 Results with USE

In this section, we focus on the response by the two hardware chipsets when running the simulation models on top of the USE platform. We still compiled USE with the -O3 flag, thus enabling the binary code optimization for performance. Also, in these experiments, we relied on the same configurations (light, medium and heavy) of both the PCS and the Highway models that have been presented in Section 5.1. Furthermore, we still report performance data (event throughput) computed as the average over 20 samples, where each run has been set to last 1 minute along wall-clock-time.

We recall that USE relies on non-blocking concurrent algorithms managing a unique fully shared event pool—in particular for arbitrating the accesses and the pick of any event buffer by every thread. These operations imply changing atomically the bits representing the state of the event buffer, plus cross-buffer pointers, which is exactly done in a non-blocking manner via RMW instructions. Furthermore, execution paths of such non-blocking algorithm could fail—for example because of the failure of a Compare-and-Swap RMW instruction—leading to the re-execution of the path according to an abort/retry protocol typical of the non-blocking paradigm. Hence, with light workload of the simulation models there is a larger incidence of RMW instructions for managing the event pool. Such an impact is expected to be greater than what happens with PARSIR, just since the access to the fully shared event pool in USE takes place systematically upon each event pick for processing. The same takes place when increasing the workload of the simulation models, although we get a reduction of the incidence of simulation engine machine instructions compared to application specific ones.

The event throughput values we observed when running with USE are reported in Figure 5 for the PCS model and in Figure 6 for the Highway model (balanced configuration). Given the speculative (optimistic) synchronization mechanism offered by USE, we report both the throughput of events that are really committed and the total throughput of events that are processed. This allowed us to determine whether the trends of the execution speed on the two hardware chipsets can really match the actual trend of useful (not rolled back) simulation work done by the simulation engine when running the models. More important, a relevant distance between throughput values for committed vs total (committed plus rolled back) events allows determining that the profile of the execution of the engine-level software of USE is even more prone to carrying out tasks, in particular memory updates, for managing the shared event queue—for example for marking still via RMW instructions the elements that need to be passed to some no longer valid execution trajectory (a kind of annihilation) hence being no longer to consider in any resume of the execution after a rollback.

The results are quite different from those we observed with PARSIR. In particular, data for the PCS model (see Figure 5) highlight how the two hardware chipsets produce very similar throughput of committed events. At the same time, the absolute execution speed (the total event throughput) is higher than the throughput of committed events. Also, the performance provided by the powerPC RISC chipset is equal, or a bit better, than the one observed with the x86 CISC chipset. This confirms the observations we made on the hardware chipsets in Section 5.1. In particular, when memory boundness increases—in this case because of the need to manage the shared event pool for annihilation of the events caused by rollbacks and also the marking of events that instead will still need to be reprocessed after a rollback—the powerPC RISC chipset provides interesting support for effective execution. Such trend is confirmed by data we gathered for the Highway model (see Figure 6). In fact, for such a model the distance between committed and total events' throughput is essentially negligible, indicating that the actual execution of the engine-level software of USE is making much less usage of memory updates for, e.g., annihilation of events. In such a scenario, the arithmetic operations by the engine (e.g., for timestamp comparison) have a larger incidence thus increasing the CPU-intensiveness of the engine-level execution. Under this setting, the x86 CISC platform provides the advantages we already discussed in Section 5.1 for the case of the PARSIR engine.

Going deeper into a few details, USE gives rise to an execution profile where the percentage of wall-clock-time spent by threads while running engine level binary code (rather than model specific one) is larger than the one of PARSIR. This is confirmed by the fact that the overall execution speed guaranteed by USE on top of both the hardware chipsets is around 1 order of magnitude lower than the one offered by PARSIR—such a result matches the analysis provided in [22], where the two PDES engines have been directly compared when running on the same (unique) chipset based on x86 processors. Such larger activity executed by USE at the simulation engine level is not only related to the execution of the support for rollback[2]. Rather, it is directly linked to the cost of the non-blocking concurrent algorithms for managing the fully shared event pool, in particular their RMW instructions and the related effects on hardware level cache-coherency/store-buffer management protocols. This leads to reduced likelihood of disjoint access parallelism, which limits the benefits provided by the powerPC RISC chipset when the model has more memory demanding behavior.

Such reduction of the benefits is noted for both the balanced Highway model (see Figure 6)—recall that this model is more memory demanding than CPU demanding compared to PCS—and also the unbalanced version (we remind the reader to Section 5.1 for the detailed description of this configuration), whose data are reported in Figure 7.

[2]Beyond the performance difference, we recall anyhow that USE can support the execution of simulation models with zero lookahead just thanks to speculative processing, which is instead not supported by PARSIR.

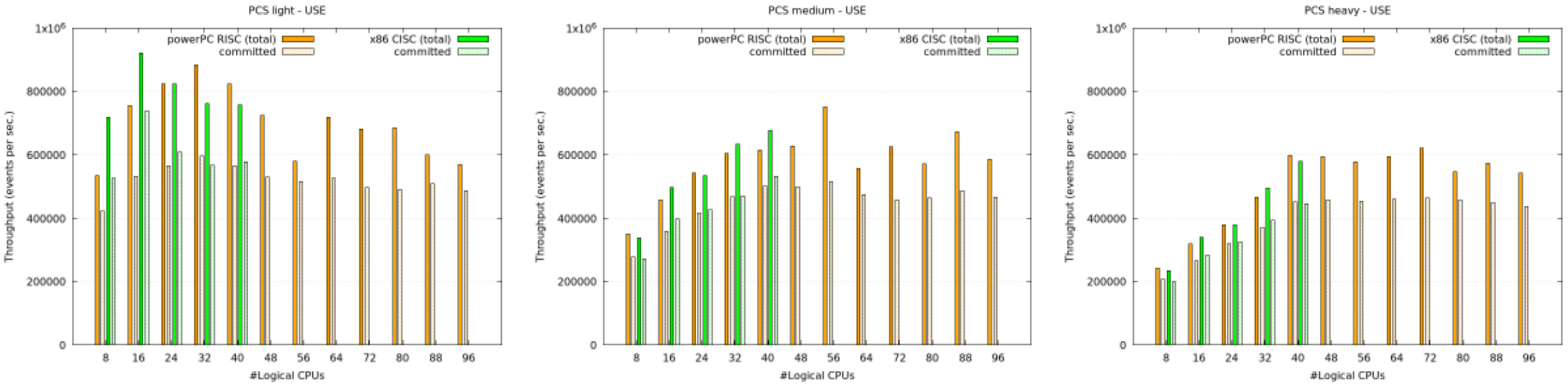

**Figure 5: Results for PCS run on top of USE.**

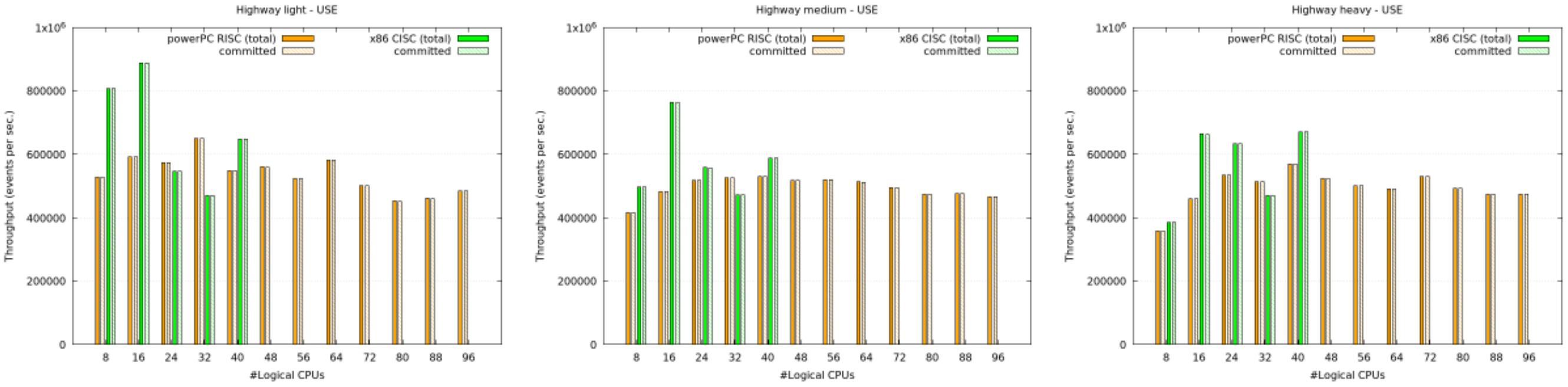

**Figure 6: Results for Highway run on top of USE.**

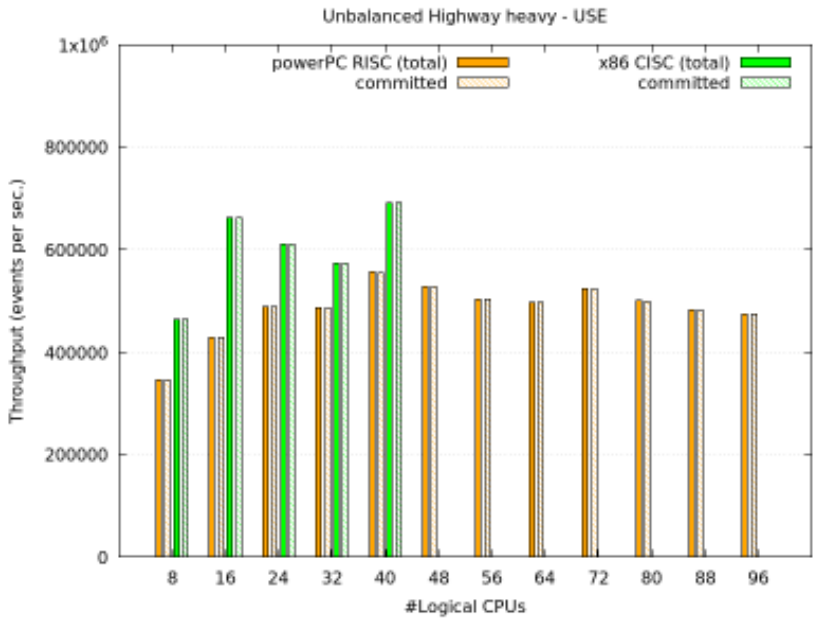

**Figure 7: Results for unbalanced Highway run on top of USE.**

## 6 DISCUSSION

The outcomes of the previous experiments allow the identification of ranges of configurations of both PDES engines, and target models to be run on top of them, for preferring one or the other of the two compared hardware chipsets. Figure 8 allows schematizing such results. In particular, we represent three-dimensional axes where, along each dimension, we report one of the three hardware-level dynamics that characterized our experiments. One (the y axis) expresses the intensity of the CPU pipeline usage by either the engine-level software or the simulation model software. The second one (the x axis) represents the impact we can expect from the usage of RMW instructions (for managing shared data structures) and their effects on cache coherency, in particular in terms of cache-line invalidation. The third one (the z axis) represents the level of disjoint access parallelism (DAP) we can expect the simulation engine provides while running both engine-level software and the overlying simulation models. We recall that such a type of access to memory leads to marginal impact on cache line invalidations, just thanks to the time-separated accesses to different memory regions by the concurrent threads.

At our abstraction level, we represent these parameters as normalized between 0 and 1. Also, to make the image well shaped, given a parameter $P$, sometimes we represent its value along the axis using $(1 - P)$—this takes place for both the y axis and the z axis.

By Figure 8, we observe the presence of a zone in the space characterized by extremely high values along the x axis (hence large/very-large impact of RMW and cache coherency, in particular related to cache-line invalidation and frequent changes of the state of cache lines in the hardware) where the two hardware chipsets appear to be somehow equivalent. At the same time, when the processor pipeline plays a core role in the execution of the PDES engine and/or the overlying model, the x86 CISC chipset appears to provide better chances of higher performance. Finally, when memory intensiveness takes place, and is in particular expressed via large incidence of DAP, the powerPC RISC chipset appears to be the best solution.

We think these deductions can be generalized, hence providing a support for selecting types of hardware vs types of PDES engines (e.g. with more or less incidence of data sharing with large volumes of concurrent accesses across threads) for running specific models (e.g. making more or less usage of CPU pipeline vs memory). Also, they still scale to scenarios where the incidence of application specific software vs PDES engine software, in terms of CPU usage by the threads, is general.

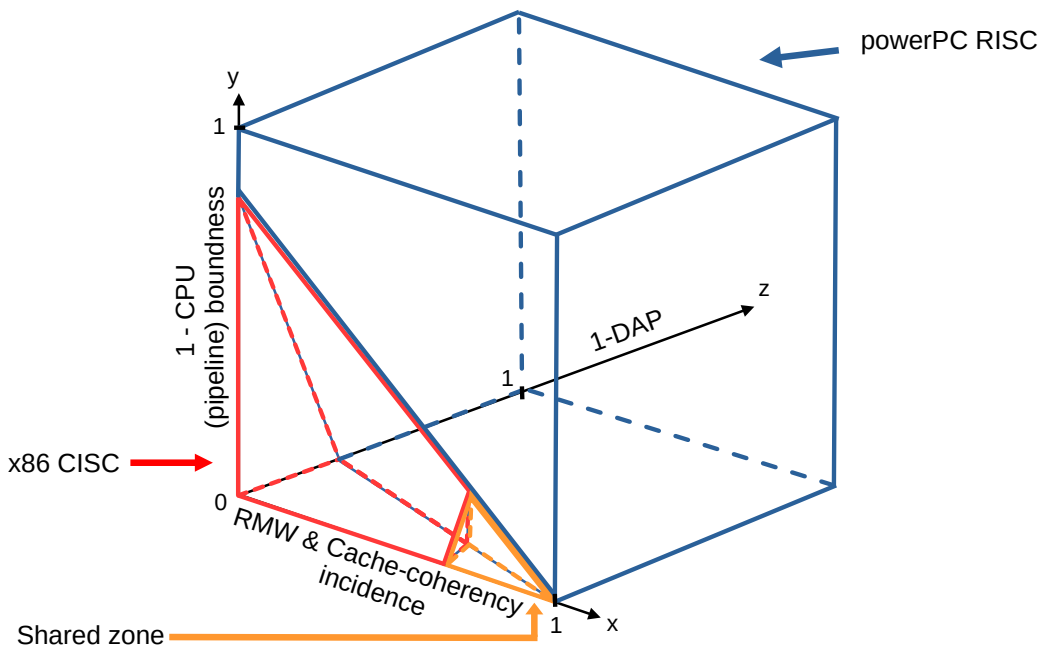

**Figure 8: Expected trends.**

## 7 RELATED WORK

In the literature, one common approach that takes into account hardware chipsets for performance and scalability optimization is software-hardware co-design. This area has been investigated in various application contexts like AI [7, 20, 35], networking [16, 36] and embedded systems [31, 34].

In other scenarios, as well as in the PDES area, heterogeneous hardware has been investigated for supporting solutions tailored for individual architectures equipped with differentiated components, like FPGA or GPU accelerators (see, e.g., [8, 24, 30, 32, 33]).

Other studies, a few of which focused on PDES, exploit specific features at the hardware level—in particular on board of common CPUs—for supporting/assessing the optimization of the thread execution flow. In this area we find proposals that rely on the Inter-Processor-Interrupt (IPI) support [18, 26] as well as on performance Monitoring Units (PMUs) [9, 25] leading, for example, to the inclusion of asynchronous switch of the thread across differentiated contexts [27].

Still focusing on the PDES area, several works have proposed investigations on how to reconfigure operating system software for managing specific hardware-related events (e.g., time-passage events) in order to improve performance and fidelity of simulation runs. Here we mostly find solutions tailored for Linux, which support virtual time-based simulation of cyber-physical systems [10] or early notification of timestamp (event) priority inversion in speculative simulation [21]. These proposals are essentially based on the exploitation of operating system API (e.g., kernel-level API), hence only indirectly focusing on the specific hardware chipset for which the actual API code has been devised.

The work in [2] provides a study where CPU vs GPU computations of agent-based simulation models are compared. Here the focus is to determine whether hardware accelerators that are nowadays commonly used can provide benefits for a specific class of simulation applications.

Overall, each of the above solutions is focused on an individual architecture, or on testing a software solution on an instance of hardware chipset, and sometimes on comparing classical CPU execution vs the outcomes achievable via accelerators (e.g., GPUs). This work is instead focused on the observation of the support offered by different, but homologue (fully CPU-based), hardware chipsets when running PDES engines—and models with specific CPU (pipeline) vs memory boundness—on top of them. The results of our experimental study are therefore orthogonal to what has already been presented in the literature.

## 8 CONCLUSIONS

This study has examined the performance of x86 (CISC) and powerPC (RISC) architectures in the context of PDES applications, evaluating two distinct PDES engines across two real-world simulation models. The results reveal that there is no definitive "winner" between the two hardware chipsets; rather, the effectiveness of each depends on the specific characteristics of the software, simulation workload, and underlying PDES engine. This can provide support for selection and investments at the hardware level.

Our findings emphasize that x86's complex instruction set and deeper pipelines may offer advantages for certain PDES engines that benefit from high in-pipeline performance along each single execution flow. Meanwhile, the streamlined instruction execution of the powerPC RISC architecture can lead to improved performance in workloads that leverage parallel execution more effectively, especially in terms of actual memory accesses. These trade-offs suggest that optimal hardware selection should be guided by the computational structure of the target simulation, rather than a one-size-fits-all approach.

The novelty of this work lies in its comparative perspective, contrasting with the prevailing focus in PDES research on optimizing a given engine for a specific chipset. By broadening the scope to evaluate multiple PDES engines across distinct simulation models, we provide deeper insights into the interplay between hardware and software in PDES applications. Future research could extend this investigation by incorporating additional architectures, hybrid computing solutions, or further exploring hardware acceleration techniques such as GPUs and FPGAs for PDES workloads.

In conclusion, while x86 and powerPC each present strengths and limitations, our results highlight the importance of a holistic approach to hardware selection in PDES, considering both architectural traits and software characteristics to achieve the best performance for a given simulation task.

## ACKNOWLEDGMENTS

The authors thank IBM Italia and Ricca IT for providing the IBM Power10 processor-based server and support to make this work possible.